\documentclass{iucrjournals}

\usepackage[capitalise]{cleveref}
\usepackage{placeins}
\usepackage{xcolor}
\usepackage[inline]{enumitem}
\usepackage{siunitx}

\title{Expanding Neutron Capacity at the ESS through Symbiotic Instruments: SLEIPNIR \& Remora}

\author[a,c]{Nicolai Lindaa Amin\IUCrOrcidlink{0009-0007-4856-2650}\IUCrEmaillink{amin@ill.fr}}%
\author[b]{Mohamed Aouane\IUCrOrcidlink{0000-0002-0129-8414}}%
\author[a,c]{Peter Kjær Willendrup\IUCrOrcidlink{0000-0002-4238-9478}}%
\author[d]{Jörg Voigt\IUCrOrcidlink{0000-0002-6556-4961}}%
\author[c]{Gregory S. Tucker\IUCrOrcidlink{0000-0002-2787-8054}}%
\author[e]{Robert Georgii\IUCrOrcidlink{0000-0002-9366-4935}}%
\author[e]{Ellen Fogh\IUCrOrcidlink{0000-0001-8305-4466}}%
\author[a]{Niels Bech Christensen\IUCrOrcidlink{0000-0001-6443-2142}}%
\author[a,b]{Rasmus Toft-Petersen\IUCrCemaillink{rasmus.toft-petersen@ess.eu}\IUCrOrcidlink{0000-0001-7638-3675}}%

\affil[a]{Department of Physics, Technical University of Denmark, Fysikvej 1, 2800 Kongens Lyngby, Denmark}%
\affil[b]{Spectroscopy Division, European Spallation Source ERIC, P.O. Box 176, SE-221 00 Lund, Sweden}%
\affil[c]{Data Management \& Scientific Computing, European Spallation Source ERIC, Asmussens Allé 305, 2800 Kongens Lyngby, Denmark}%
\affil[d]{Jülich Centre for Neutron Science, Wilhelm-Johnen-Straße, 52428 Jülich, Germany}%
\affil[e]{Heinz Maier-Leibnitz Zentrum, Technical University of Munich, Lichtenbergstr. 1, 85748 Garching, Germany}%
\begin{document} 
\maketitle 

\begin{synopsis}
We introduce two novel symbiotic neutron instrument concepts that exploit unused spatial and spectral capacity at the European Spallation Source, enabling additional instruments without requiring new beam ports. Monte Carlo simulations demonstrate that both concepts achieve performance comparable to existing productive instruments at other neutron sources
\end{synopsis}

\begin{abstract}
The long pulse of the European Spallation Source (ESS) offers unique opportunities to host more than one neutron instrument on a single beam-port. Here, we present two symbiotic instrument concepts designed for previously unexploited neutrons, without imposing significant perturbation on existing or planned beamlines. The first concept employs a doubly focusing dual-monochromator system to vertically translate the neutron beam, enabling compact instruments to be stacked within spatially constrained beam ports inaccessible to conventional instruments. The second extracts excess spectral bandwidth from an existing guide via Bragg reflection upstream of the flagship instrument's chopper system, feeding a second instrument. Both concepts are optimized using gradient-descent Monte Carlo ray-tracing simulations in McStas and benchmarked against existing productive instruments at other sources. Despite operating under significant geometric and flux constraints, the resulting instruments demonstrate performance comparable to well-established reference instruments at the ILL. The two concepts together illustrate a modular and cost-effective route to expanding ESS scientific capacity, using the facility's resources that would otherwise go unused.
\end{abstract}


\section{Introduction}\label{Intro}
The neutron facility landscape has contracted over the last few decades: where numerous medium-brilliance sources once supplemented a handful of flagship facilities, routine experiments, method development and training now compete for time at the same few large-scale sources, restricting beamtime availability across the community. Several new facilities and smaller university sources have been proposed to address this capacity challenge \cite{Carpenter2019}.

The European Spallation Source (ESS) is slated to become the world's brightest neutron source once full 2 MW operation is reached. Its suite of 15 flagship instruments, nearing completion, is designed chiefly to expand scientific capability, with an ongoing road map process now identifying the next generation. Yet the ESS is also uniquely positioned to expand capacity: as a long-pulse source with unusually high integrated brilliance and 42 beam ports, it offers more usable neutrons than its flagship suite alone can exploit. Many instruments extend beyond 150~m, leaving open space between beamlines far from the source, while near the target the shielding footprint of high-flux instruments renders several beam ports inaccessible to further large-scale instruments altogether.

Here, we present two symbiotic instrument concepts that use neutrons otherwise lost to the flagship programme, while remaining minimally invasive to the ESS's primary mission of world-leading science through that programme: i) a vertically stacked double-bounce monochromator system feeding several smaller instruments, including single-crystal and powder diffractometers, and ii) a crystal-monochromator direct-geometry spectrometer exploiting excess bandwidth from the long ESS pulse. Neither concept is expected to match flagship performance, but as we show below, both are comparable to highly productive instruments at other facilities. Scientifically relevant beamlines can therefore be realized at comparatively low cost and without major new infrastructure, by exploiting the ESS's brilliance, pulse structure and existing spatial opportunities.


\section{Spatial Constraints and Opportunities}\label{sec:Design_overview}

\begin{figure}[h]
    \begin{center}
    \includegraphics[width=0.75\textwidth]{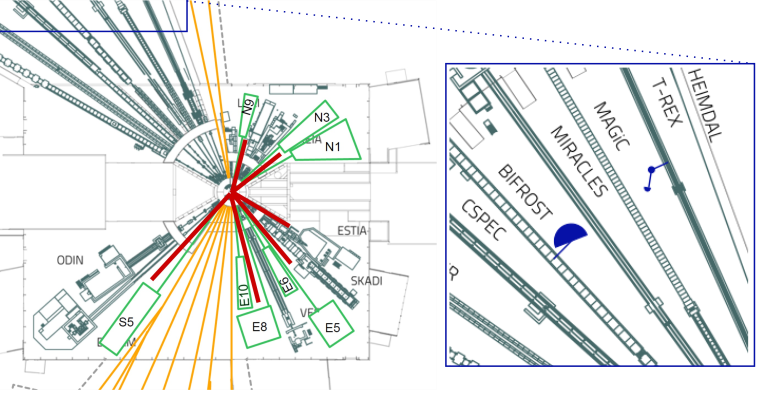}
    \caption{Left: Schematic of the proposed expansion of the ESS instrument hall, with red beamlines highlighting the spatially constrained beam ports N2, N8, E1, E4, E9, and S4, which cannot accommodate instruments with large spatial footprints due to the neighboring footprints of other already-built (grey) or conceptual (green and orange) instruments and beamlines. The blue box shows the long guide hall in the West Sector. Inset: Schematic of possible positions for symbiotic instruments between the long ESS beamlines in the West sector.}
    \label{fig:blocked_beamlines}  
    \end{center}
\end{figure}

Of the 42 beam ports available at the ESS, only 16 will be utilized by 2027, corresponding to a utilization of 38\%. Full utilization of all beam ports for flagship instruments appears infeasible, as the spatial constraints imposed by the shielding geometry preclude the construction of large-scale instruments at several beam ports. The N2, N8, E1, E4, E9 and S4 beam ports, highlighted in \cref{fig:blocked_beamlines}, represent such cases. The shielding requirements of the large high-flux instruments, and the space requirements of their choppers close to the target, leave little room for large instruments at those positions, and therefore they are not currently considered for such instruments, representing a loss of access to ESS brilliance.

\begin{figure}[h]
	\centering
	\includegraphics[width=\textwidth]{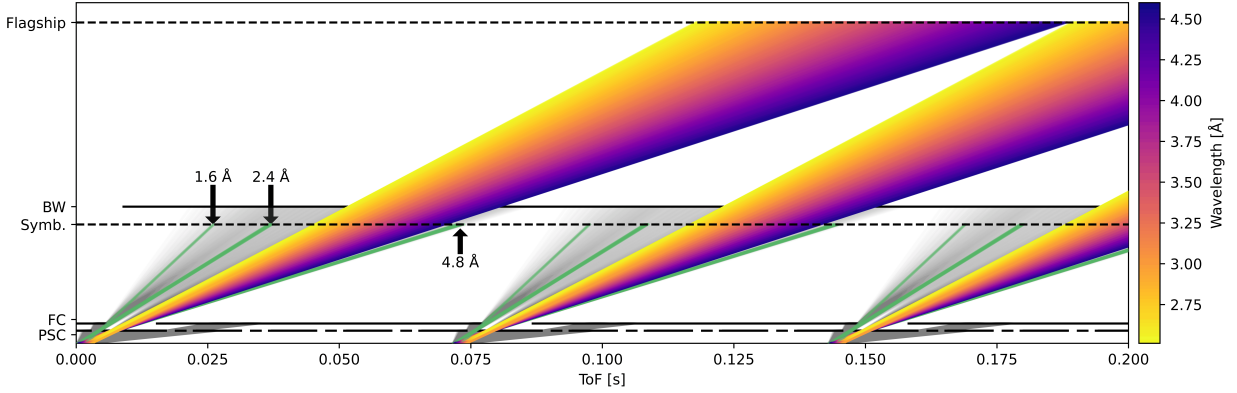}
	\caption{Time-of-flight diagram illustrating a proposed chopper cascade for the symbiotic spectrometer. The bandwidth chopper (BWC) divides the pulse into two operational windows (indicated by the green lines): prior to closure, the first and second harmonics are transmitted to the symbiotic instrument, while the fundamental wavelength at \SI{4.8}{\angstrom} is utilized after closure. The wavelength band of the flagship instrument remains largely unaffected by the upstream monochromator extraction. Grey lines indicate neutrons that are unused by either flagship or symbiotic instruments. The neutron beam is shaped by the PSC and overlap between pulses is avoided by the Frame overlap Chopper (FC). Adapted from \protect\cite{Fogh2026ProposalRemora}.}
	\label{fig:Remora_ToF}
\end{figure}

The ESS long-pulse structure delivers a usable pulse width of \SI{2.86}{ms}, yet instruments employing pulse-shaping choppers (PSCs) routinely deliver only a fraction of the available neutron bandwidth to the sample. Spectrally, the ESS moderators spans a usable wavelength range from approximately \SI{0.5}{\angstrom} to \SI{20}{\angstrom}, yet long instruments are constrained to exploit only a narrow band. Long instruments in the West sector, as indicated in \cref{fig:blocked_beamlines}, operate over approximately \SI{1.7}{\angstrom} of bandwidth under standard conditions, this limit is set by the length of the instruments and the position of the PSC: With an infinitesimal PSC opening time, only the desired bandwidth is accepted by the chopper cascade, with frame overlap occurring at 162 meters \cite{Garoby2018TheDesign}. However, in high flux settings, with a large PSC opening time, a significantly larger bandwidth is accepted. In such cases, frame overlap is avoided by using a bandwidth chopper (BWC), which limits the bandwidth halfway downstream the primary spectrometer, see \cref{fig:Remora_ToF}. The excess bandwidth is available for symbiotic use before the BWC, at limited performance cost for the flagship instrument. 

The spatial arrangement of the ESS instrument hall presents a natural opportunity to recover a portion of this otherwise unused flux. With an angular separation of approximately \SI{6}{\degree} between adjacent long beamlines \cite{Andersen2020TheSource}, the lateral separation between guide systems at \SI{50}{m} from the moderator reaches approximately \SI{4}{m}, assuming \SI{0.5}{m} of shielding on each side of a guide. This inter-guide space is sufficient to accommodate secondary extraction optics, a small shielding enclosure, and a compact instrument. \cref{fig:blocked_beamlines} Inset illustrates 2 candidate locations: one between BIFROST and MIRACLES, fed by the BIFROST guide, and one between MAGIC and T-REX, fed by the T-REX guide.


\section{Methodology}\label{sec:Methodology}

The design optimization of instrument concepts in these limited spaces was carried out using the Monte Carlo ray-tracing package McStas \cite{Lefmann1999McStasSimulations, Willendrup2004McStasPackage}. McStas provides detailed component models for ESS neutron instruments, including a validated source description, which served as the common starting point for both designs.

\begin{figure}[h]
	\centering
	\includegraphics[width=\textwidth]{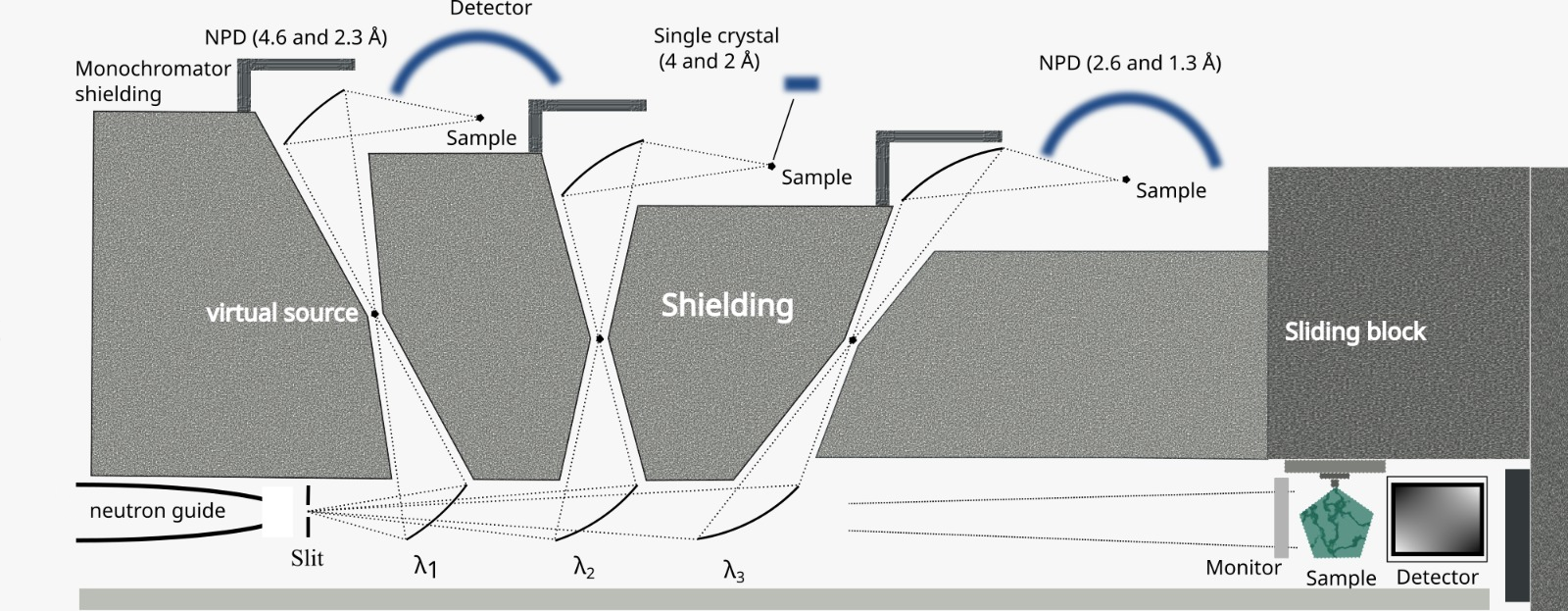}
	\caption{Possible configuration of a SLEIPNIR suite, with four instruments: two powder diffractometer and a single crystal diffractometer, which are vertically translated, and an imaging station that utilizes the last part of the floor space. Each of the vertically translated instruments utilize different wavelengths and their harmonics. The virtual sources (black dots between sets of monochromators) is covered by shielding to minimize crosstalk between the different instruments.}
	\label{fig:SLEIPNIR_Total}
\end{figure}

A gradient descent algorithm was employed to navigate the multi-dimensional parameter spaces of each concept. The procedure operates as follows: a third-degree polynomial is fitted to the figure of merit (FOM) as a function of each free parameter (e.g., component distances, mosaicities, or collimation angles) in turn. The fitted FOM provides an estimate of the local gradient; min–max normalization, rescaling each parameter and the FOM to the range [0,1] based on their sampled minimum and maximum values, is applied to ensure a uniform effective step size across all dimensions, and effectively removing any units from the parameters or FOM; the parameters are then updated iteratively in the direction of increasing FOM until no further improvement is observed over several consecutive cycles.
The free parameters encompassed distances between optical elements, component dimensions, monochromator mosaicities, and collimation settings. A fixed step size of $\eta = 0.1$ was used throughout the optimization runs \cite{ruder2017overviewgradientdescentoptimization}. The principal limitation of gradient descent, susceptibility to local maxima, was mitigated by polynomial smoothing of the noisy simulation data and by systematic monitoring of convergence behavior across multiple independent starting points. Running the optimization from several starting points and comparing the resulting plateaus serves as a practical, if not exhaustive, check against convergence to a poor local optimum. Each free parameter was scanned over 40 linearly spaced values with $10^7$ simulated neutrons per setting, and the iterative procedure was continued until the FOM reached a stable plateau. After finding the plateau a longer simulation was run, such that simulated flux values have a relative uncertainty of $\leq 0.5 \%$.

\section{SLEIPNIR - Vertically Translated Instruments}\label{sec:Design-SLEIPNIR}

The first symbiotic concept, hereafter referred to as SLEIPNIR after Odin's eight-legged horse, in reference to the multiple stacked \emph{legs} of the design, addresses the class of spatially constrained beam ports described in the first part of \cref{sec:Design_overview}. The concept relies on a doubly focused dual-monochromator system that vertically translates the monochromatized neutron beam. The first monochromator, on the instrument hall nominal level, focuses  at a vertical angle onto a virtual source encased in shielding; while the second employs point-to-point focusing at the same takeoff angle, resulting in an elevated beam, parallel to the hall floor. Multiple independent monochromatic \emph{legs} can be stacked within a single confined footprint, each operating at a distinct wavelength. The long-pulse ESS source further allows both the fundamental wavelength and its second harmonic to be exploited within a single leg, discriminated by monochromators rather than choppers, provided sufficient source-to-sample distance.

The concept is modular: legs that accommodate two powder diffractometers, a single-crystal diffractometer, and an imaging station could in principle be operated simultaneously from a single beam port, as illustrated in \cref{fig:SLEIPNIR_Total}. The principal trade-off is a flux reduction of at least one order of magnitude relative to flagship ESS instruments, arising from using two monochromators in series. However, the high brilliance of the ESS source mitigates this reduction significantly, as for a broad class of experiments the delivered flux is sufficient. The proposed imaging station at the end of the guide will receive the transmitted neutrons, with only limited loss of wavelength bandwidth due to the upstream instruments.

For the purposes of this study, we focus on a single leg implemented as a powder diffractometer, which is optimized and compared to the D1B diffractometer at the ILL \cite{Orench2014TheLangevin}, a well-established and productive reference instrument. Co-optimization of multiple legs is a topic for future studies. The powder diffractometer is modeled in McStas beginning with an elliptical guide that transports flux from the ESS moderator to a vertically translating monochromator pair. A collimator defines the beam divergence both vertically and horizontally after the second monochromator and before the sample position, and a position-sensitive detector bank covers a continuous range of scattering angles around the sample, $5^\circ$-$130^\circ$. Monochromator curvatures are calculated analytically from the relevant source to monochromator and monochromator to focal point distances via Rowland geometry \cite{BROHOLM1996169}, fx. virtual source to second monochromator and second monochromator to sample, rather than treated as free variables. The fundamental wavelength is fixed at \SI{2.6}{\angstrom} to allow direct comparison with established powder diffractometers operating at similar wavelengths, with monochromator reflectivities of 0.9 and 0.3 for the fundamental and first harmonic respectively \cite{Freund2022OptimizationOptics,Moore2000NeutronGraphite, Adib2012CharacteristicsMonochromator}. The distance from moderator to the end of the guide is set to the minimum permitted by the distance of the ESS bunker. All remaining parameters, guide dimensions, monochromator mosaicities, distances between optical components, and collimation settings, are free variables. The only limitation set is that the height of the instrument is \SI{3.5}{m}, not including the detector bank. This limitation is set to avoid the structure becoming unmanageable and potentially interfering with existing structures, and infrastructure associated with other instruments. The instrument was optimized using the FOM of Cussen et al. \cite{Cussen2016ComputerConfigurations} as a well-defined numerical basis for jointly evaluating resolution and flux.

\subsection{Results}\label{sec:SLEIPNIR_Results}
\begin{figure}[h]
    \centering
    \includegraphics[width=\textwidth]{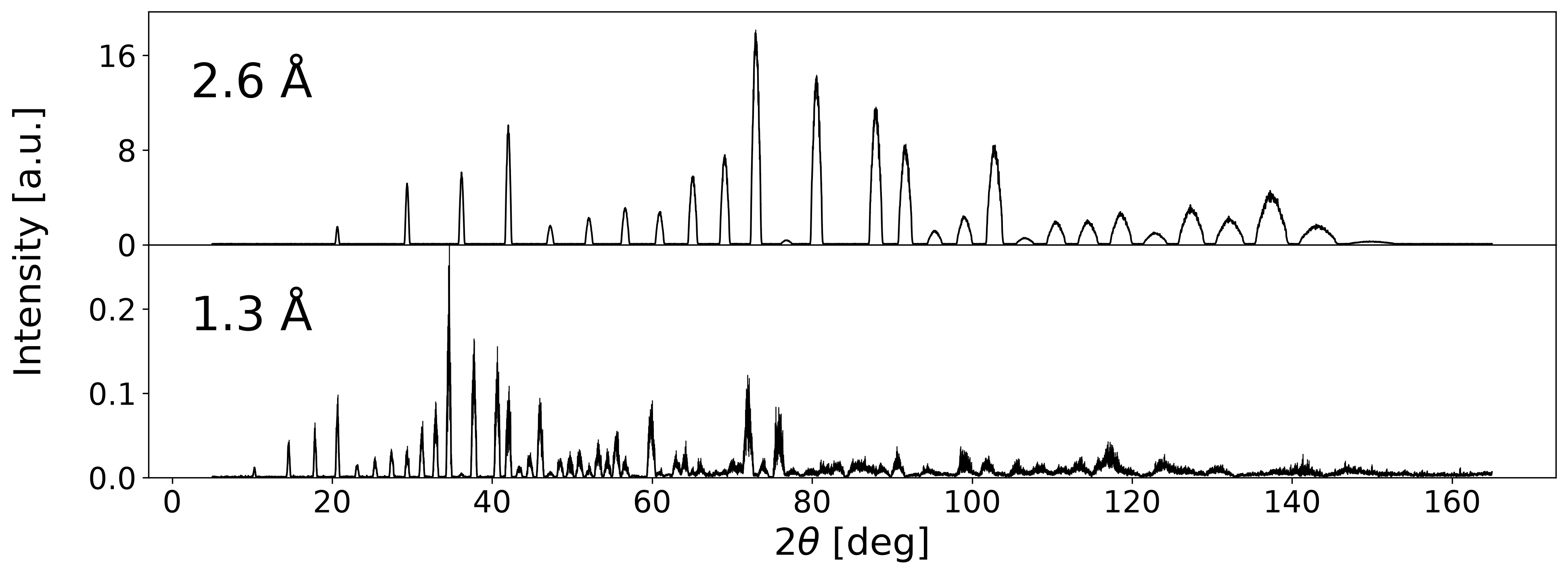}
    \caption{Simulated diffraction spectra of a $\text{Na}_2\text{Ca}_3\text{Al}_2\text{F}_{14}$ sample for the optimized powder diffractometer for the full (top) and half (bottom) wavelength.}
    \label{fig:SLEIPNIR_Res_NAC}
\end{figure}

The gradient-descent optimization converged to a single, robust design, balancing flux and angular resolution within the geometric constraints imposed by the vertical translation scheme. The resolution of the instrument is largely governed by the initial monochromatization of the beam by the first monochromator, where the wavelength distribution is effectively defined. Downstream optics primarily act to focus as many neutrons as possible onto the sample, contributing only marginally to further narrowing of the resolution. The large distances between the monochromator and sample position allows for distance collimation of the beam, which improves the resolution, but lowers the flux, and makes the beam spot at the sample position larger. This places the instrument naturally in the medium-resolution regime. The specific instrument parameters and McStas files are available in \cref{sec:SLEIPNIR_params} for repeatability.

\begin{figure}[h]
    \centering
    \includegraphics[width=0.75\textwidth]{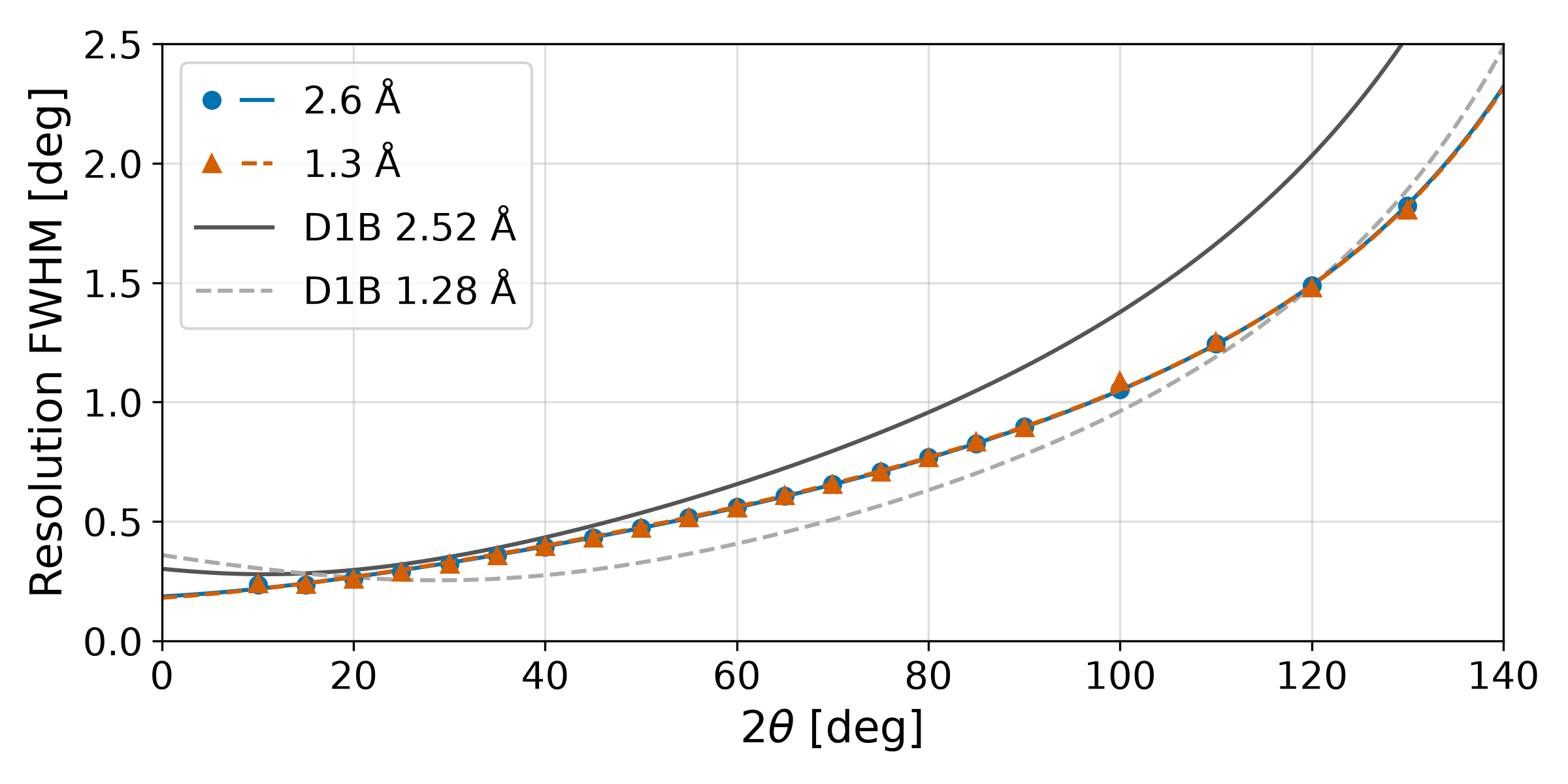}
    \caption{Resolution functions of the powder diffractometer, fundamental (blue) and 1st harmonic (orange), as well as D1B at high (dark grey) and low (light grey) resolution setting. The fits are made using the Cagliotti formulation for powder diffractometer resolutions \protect\cite{Caglioti1962ResolutionDiffraction}}
    \label{fig:SLEIPNIR_Res}
\end{figure}

\begin{table}[ht]
\caption{Comparison of SLEIPNIR and D1B flux at the sample position and resolution-function parameters. D1B values were obtained from the ILL website.}
\label{tab:PD_comp}
\begin{center}
\begin{tabular}{lcccc}
\midrule
Instrument & Flux at S.P. & $U$ & $V$ & $W$\\
\midrule
SLEIPNIR (2.6 \AA) & $3.1\cdot10^6$ & 0.677 & 0.089  & 0.035 \\
SLEIPNIR (1.3 \AA) & $3.5\cdot10^4$ & 0.667 & 0.104  & 0.033 \\
D1B (2.52 \AA)     & $7.9\cdot10^6$ & 1.508 & $-0.283$ & 0.091 \\
D1B (1.28 \AA)     & $4\cdot10^5$   & 0.984 & $-0.505$ & 0.129 \\
\midrule
\end{tabular}
\end{center}
\end{table}

The comparison of performance to D1B at the ILL is given in \cref{tab:PD_comp} and \cref{fig:SLEIPNIR_Res}. At the fundamental wavelength, \SI{2.6}{\angstrom}, the stacked diffractometer achieves a sample flux of approximately 38\% of D1B at its high-intensity setting. Importantly, the Caglioti resolution parameters \cite{Caglioti1958ChoiceDiffractionb, Caglioti1962ResolutionDiffraction} and \cref{fig:SLEIPNIR_Res} show a significantly flatter resolution curve across the full angular range, with the stacked diffractometer having slightly finer resolution in the backscattering region. This is a direct consequence of the tighter wavelength definition imposed by the first monochromator, due to it's low mosaicity, and the large distance collation of the second monochromator. Similarly, comparing to other powder diffractometers, the SLEIPNIR powder diffractometer performs comparably to the closed reactor BER-II diffractometers E6 and E9, with a better resolution than the former and a higher flux than the latter \cite{E6HZB,Franz2017E9:II}. These instruments were highly subscribed when BER-II was still operational, furthermore, E9 is being moved to FRM II to increase capacity. Other currently operational highly subscribed instruments such as BT-1 at NIST\cite{HighNIST}, or G41 at the closed Orphée reactor \cite{Ott2018PerformancesSource} perform similarly to the proposed SLEIPNIR powder diffractometer, with fluxes in the order of $10^6$ \si{n/s/cm^2}.

The second harmonic, arising naturally from the long-pulse ESS source structure, provides access to a shorter wavelength at reduced flux, as shown in \cref{fig:SLEIPNIR_Res_NAC}. While not a primary design target, this represents a useful byproduct that extends the accessible d-spacing range without additional instrument complexity. This is an advantage over conventional monochromator-based diffractometers, which require separate configurations for different wavelengths to filter out higher-order contributions. Another advantage is the possibility to discriminate fast neutron background via time-of-flight, improving the signal-to-noise ratio. However, the introduction of a time element also increases the complexity of the detector systems, which must be capable of resolving the arrival time of neutrons—a requirement not present in constant-wavelength instruments. In any case, the flux and resolution for both wavelengths are high enough to deliver the science of contemporary user programs and the standard instrumentation renders data analysis relatively straightforward.    

\subsection{Discussion}\label{sec:SLEIPNIR_Disc}
The height limitation forced the monochromator distances to remain coupled, restricting both flux and resolution. Lifting the height limit would allow substantially more flexible instruments. Another limitation is the choice of monochromator material. Only HOPG delivers high reflectivity for selected wavelengths and high transmission for off-Bragg scattering neutrons. Using copper limits the number of instruments possible on a single beamline due to beam attenuation, and using thin silicon wafers introduces very high losses via the low reflectivity of a two-monochromator system. 

The cost profile of this concept is very favorable, compared to the flagship instruments. The optical chain consists entirely of a single shared guide, standard crystal monochromators, collimators, and a position-sensitive detector. These components are well-established, commercially available, and require no novel infrastructure. Most importantly, this capacity expansion does not come at the expense of a future capability-expanding flagship instrument. Ensuring that such capacity expansion does not compromise future large-scale instruments is essential, both for the scientific success of the flagship instrument and for maximizing the overall impact of ESS as a facility. Furthermore, a SLEIPNIR powder diffractometer and the current state-of-the-art powder diffractometer at the ESS (DREAM \cite{Schweika2016DREAM-aESS}) are fully complementary: by handling standard measurements, inherently slow parametric studies and method development, SLEIPNIR acts to support the flagship user program and its associated user community.

Despite restrictions, the performance of the optimized diffractometer model remains competitive with medium-flux diffractometers, demonstrating that the ESS brilliance enables even a compact, constrained instrument to deliver powder diffraction data suitable for user program science. Perhaps most significantly, the powder diffractometer studied here represents only one realization of a broader concept. A single crystal diffractometer could be added in a similar way, and individual legs can be optimized for high resolution or high flux measurements. The modular nature of the concept also supports future expansion.  
The science case for medium-flux powder diffraction instruments remains strong, especially when sample environment parameters or physical processes vary on a longer time-scale than typical measurement times, which can be the case when studying Li-ion batteries\cite{nazer_operando_2016}  , hydrogen storage materials\cite{nazer_operando_2017} or in-situ magnetic nanocomposite synthesis\cite{ahlburg_exploring_2020}. With single crystal neutron diffraction, systems with large magnetic moments are easily studied on medium-flux instruments \cite{Kenzelmann} - especially when fast neutron background can be supressed. If 3 diffraction legs can be constructed, it would offer the user community an additional 500 days of beamtime yearly, enabling several dozens of additional experiments, at a cost similar to a single additional flagship instrument.

\section{Remora - Symbiotic Instruments}\label{sec:REMORA-Design}

The symbiotic spectrometer concept, Remora, exploits a different class of untapped resource: the spectral bandwidth transmitted by existing ESS guides that is not consumed by the flagship instrument. Rather than occupying a new beam port, a monochromator is inserted into an existing guide upstream of the flagship instrument's final chopper system, extracting a fraction of the beam via Bragg reflection and redirecting it toward a co-located secondary instrument. Due to the near-transparency of HOPG off the Bragg condition, the primary transmission is largely preserved, and both instruments operate in parallel from a single guide. This strategy is already well established at continuous reactor sources such as the ILL and PSI, and is here adapted to the long-pulse spallation environment of the ESS.

The monochromator extraction introduces a set of technical constraints that must be managed carefully. As illustrated in \cref{fig:Remora_ToF}, the wavelength band selected for the secondary instrument must not overlap with the operational requirements of the flagship, and the insertion of any optical element inevitably modifies the transmitted beam through absorption and incoherent scattering contributions. However, for long instruments, where the post-monochromator flightpath can reach 80-100 m, such effects can have little impact on the flagship beam profile. Higher-order harmonics scattered by the monochromator causes dips in the flux distribution, though their impact is substantially smaller than that of the primary reflection. With careful selection of monochromator material and geometry, the disturbance to the flagship instrument can in principle be minimized; confirming this quantitatively for a specific host such as T-REX requires dedicated simulation studies, which we identify as necessary future work, when suitable host instruments are being considered.
\begin{figure}
    \centering
    \includegraphics[width=\textwidth]{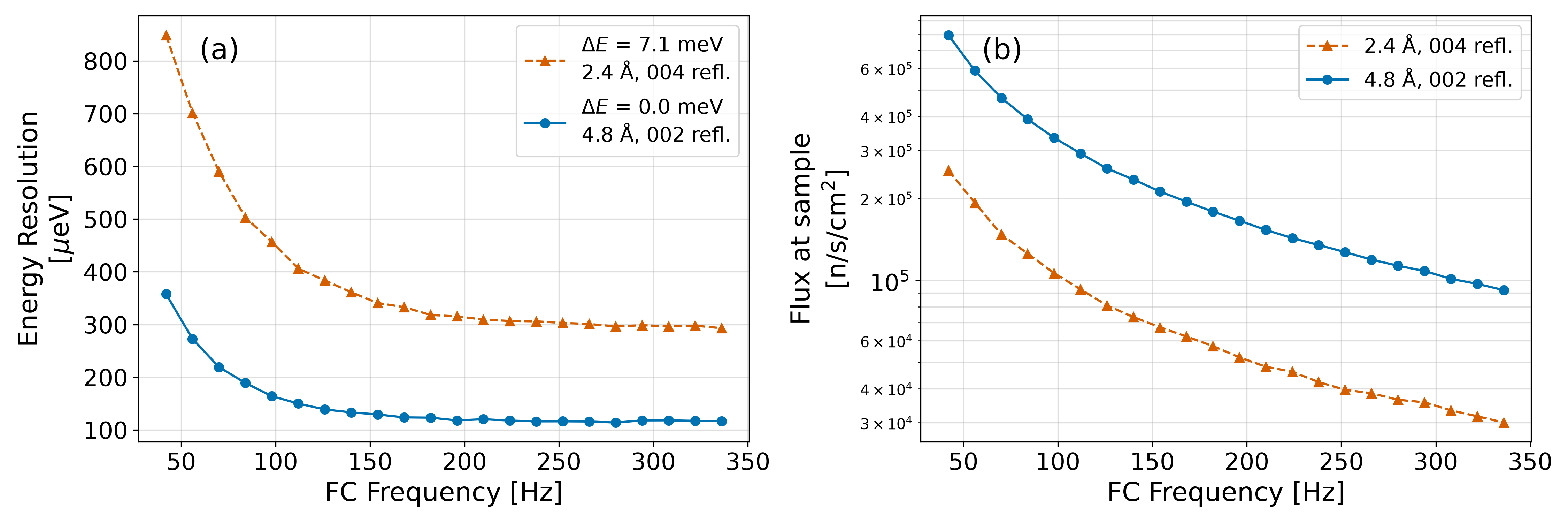}
    \caption{(a) Energy Resolution for the fundamental wavelength's elastic scattering (blue) and first harmonics inelastic 50\% energy transfer (orange), highest energy transfer of a single wavelength, of REMORA for different Fermi Chopper frequencies. (b) Flux of the fundamental (blue) and harmonic (orange) wavelengths.}
    \label{fig:REMORA_Res}
\end{figure}

The ESS long-pulse structure introduces an important advantage for symbiotic instruments compared to reactor sources or short-pulse spallation sources. The pulsed time-of-flight framework naturally separates successive harmonic reflections in arrival time at the detector, meaning that the HOPG (002) and (004) reflections of a single monochromator define two distinct, non-overlapping incident energy components within the same instrument geometry.This provides a compact route to accessing two well-separated incident energies via distinct harmonics, and thus an extended energy-transfer range, without additional optical complexity.

For the present study, the secondary instrument is implemented as a direct-geometry crystal-monochromator spectrometer, conceptually analogous to established reactor-based instruments such as PANTHER or SHARPER at the ILL \cite{Fak2022PantherILL, Zanotti2026TheSHARP}. The instrument is modeled in McStas using the T-REX guide system up to the extraction point \cite{Voigt2014ChopperSources}, due to the guides large phase-space. Here, a HOPG vertically focusing, horizontally flat monochromator reflects the chosen wavelength band, with a fundamental wavelength of \SI{4.8}{\angstrom}, toward the secondary instrument. The same reflectivity was used for the fundamental and harmonic reflections as in \cref{sec:Design-SLEIPNIR}. The extracted beam is collimated before being focused horizontally by a nested mirror optics (NMO) assembly \cite{Herb2022NestedFocusing}, transporting flux efficiently to the sample position. A single Fermi Chopper (FC), modeled as a scaled up version of the FC on SHARPER at the ILL \cite{Zanotti2026TheSHARP}, placed downstream of the monochromator and directly upstream of the sample, controls the illumination time and minimizes the time spread arising from the monochromator's finite wavelength band. A \SI{2.5}{m} sample to detector distance is used. The parameters (moderator to monochromator distance, monochromator mosaicity, NMO length and position and FC position) are free variables in the optimization. The number of NMO sections is limited to 10, to minimize complexity and production cost. The optimization targets two end goals: maximizing monochromatic flux at the sample position while maintaining an energy resolution of $\Delta E / E  \leq 3 \%$ at the elastic line for the highest FC frequency. This threshold ensures scientifically useful energy resolution without sacrificing flux to an extent that makes the instrument untenable. 
\begin{figure}[h]
    \centering
    \includegraphics[width=\linewidth]{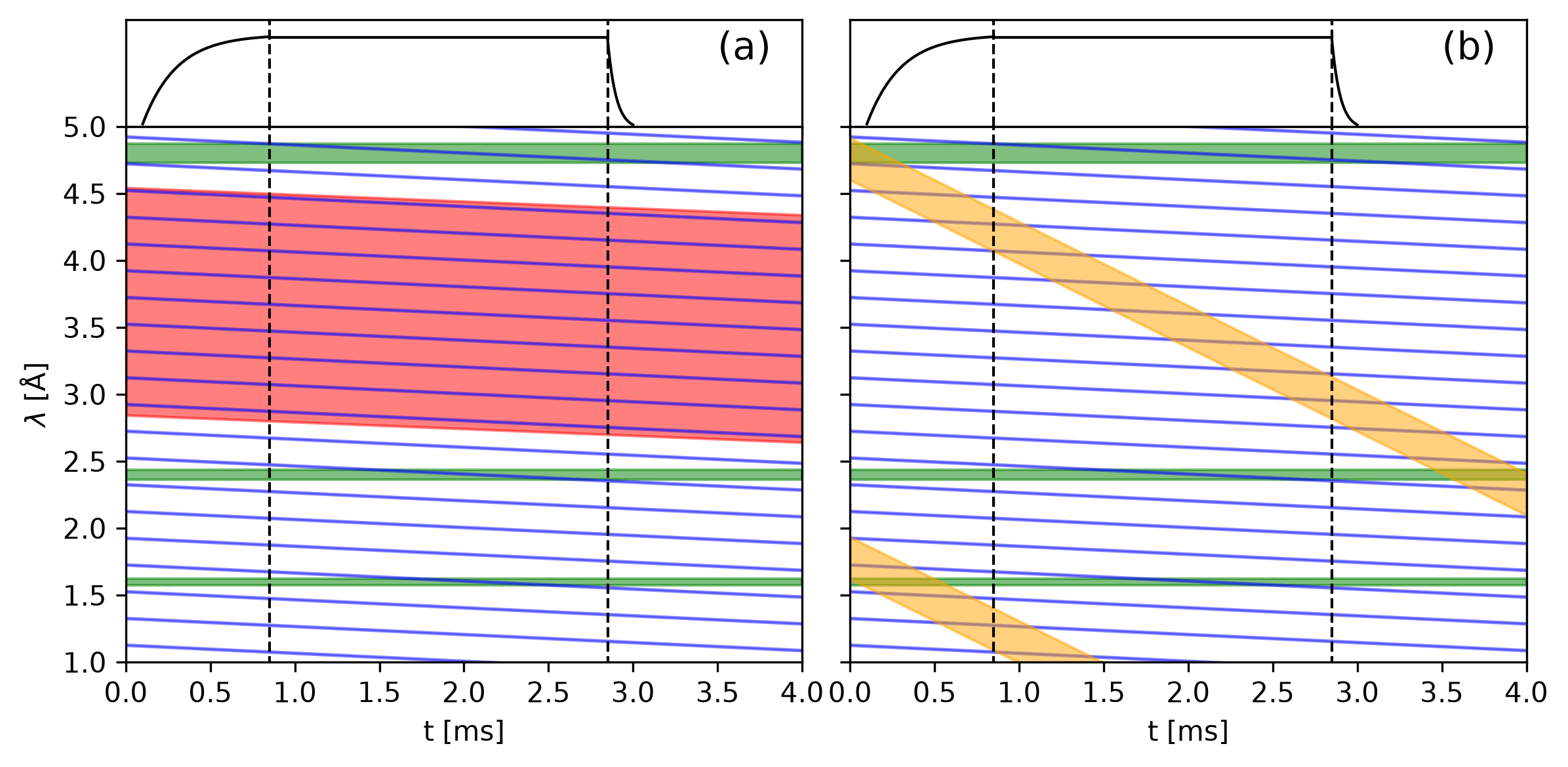}
    \caption{Acceptance diagrams of the REMORA spectrometer. Top figures of (a) and (b) show the ESS pulse, with dashed lines indicating a \SI{2}{ms} interval where the pulse is at max. (a) shows the BW chopper (red), monochromators accepted wavelength interval (green) and Fermi Chopper (blue) acceptance diagrams. There is no overlap between the BW and monochromators, as intended, as the symbiote should not impact the flagship instrument. (b) shows the PSC chopper (yellow) with an opening time of 0.5 ms for the same wavelength interval as the BW chopper in (a), monochromators accepted wavelength interval (green) and Fermi Chopper (blue). There is little overlap between the PSC chopper and the accepeted symbiotic wavelengths.}
    \label{fig:REMORA_acc}
\end{figure}

\subsection{Results}\label{sec:REMORA_Results}
Unlike the stacked diffractometer, the gradient-descent optimization of the symbiotic spectrometer did not converge to a single robust solution, but instead revealed a landscape of multiple local optima with distinct flux-resolution trade-offs. This is a direct consequence of having a crystal direct geometry spectrometer on a long pulsed source: without a pulse-shaping chopper, the energy resolution is governed primarily by the monochromatic pulse width delivered by the crystal monochromator. The competing demands of flux and resolution pull the optimization in different directions depending on the weighting applied. A representative configuration was selected as a practical compromise, achieving a resolution close to the 3\% elastic resolution target at a FC frequency of \SI{300}{Hz}, with a flux of $10^5$~\si{n/s/cm^2}, on a sample cross-section of \SI{2.2}{cm}$\times$\SI{1.9}{cm}, which is comparable to the flux delivered by LET at ISIS under similar resolution conditions (see \cref{sec:REMORA_Disc}). The dependence of energy resolution and flux on FC frequency across this solution space is shown in \cref{fig:REMORA_Res} for both the fundamental wavelength at elastic scattering and the harmonic at 50\% energy transfer. The specific instrument parameters and McStas files are available in \cref{sec:REMORA_params} for reproducability.

\begin{figure}[h]
    \centering
    \includegraphics[width=\linewidth]{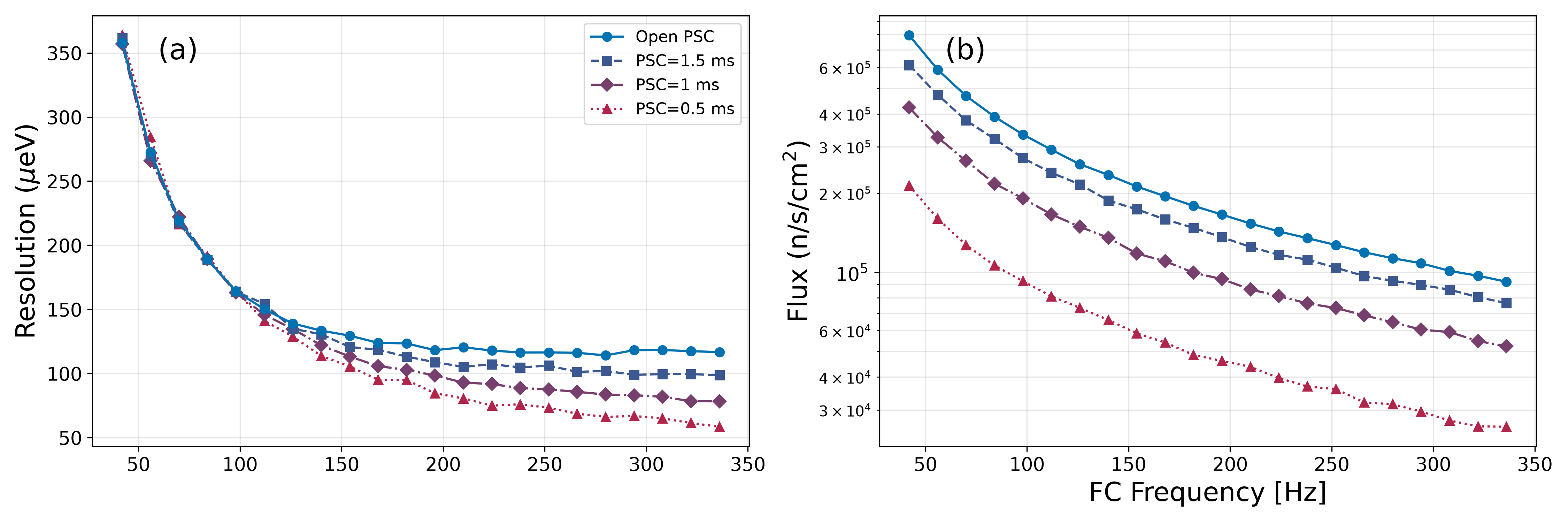}
    \caption{REMORA 4.8 Å (a) Elastic Resolution and (b) Flux as a function of Fermi Chopper speed for different PSC opening times.}
    \label{fig:PSC_REMORA}
\end{figure}

Several features are worth noting. Above approximately 150 Hz FC frequency, further increases in chopper frequency yield no resolution improvements while flux continues to fall, indicating that the instrument has entered a regime where resolution is dominated by the long source pulse rather than the FC opening time. This sets a natural operating optimum at or below this frequency. The absence of a pulse-shaping chopper, while limiting for the resolution, is a deliberate design choice rather than an oversight. As illustrated in \cref{fig:PSC_REMORA}, introducing one would sharpen the resolution of the symbiotic instrument, as expected from the inherent trade-off between resolution and flux in such an instrument \cite{Mutka1994CoupledScattering, Janssen2000TimeSpectrometer}. However, because the PSC reduces the ESS pulse available to the downstream flagship instrument, high-resolution operation of the symbiote would naturally be coordinated with the operating conditions of the host instrument. The PSC design must also preserve the relevant spectral components: if the pulse is shortened too aggressively, the fundamental or harmonic wavelength, or both, may be partially truncated. This constraint could be addressed through a variable monochromator, although this would introduce additional components into the beam and consequently affect the flux delivered to the flagship instrument. Alternatively, more sophisticated PSC designs could provide multiple openings, enabling flexible pulse shaping while retaining compatibility with the operating requirements of the flagship instrument.

The tunability of the flux-resolution trade-off extends beyond choppers. 
The focusing optics provide a continuously tunable flux-resolution trade-off: moving the nested mirror optics closer to the sample increases flux at the cost of beam divergence and resolution, while moving them further away narrows the divergence and sharpens the resolution. The same optical architecture can therefore be configured for either high-flux or high-resolution operation by adjusting component positions alone, without structural changes to the instrument. This flexibility is particularly valuable in a symbiotic context, where the scientific priorities of the secondary instrument may evolve independently of the host.

\subsection{Discussion}\label{sec:REMORA_Disc}
As a point of comparison, LET at ISIS achieves a flux of $5\times10^4$~\si{n/s/cm^2} at 2\% elastic energy resolution for a single similar wavelengths~\cite{Bewley2011LETISIS}, while the symbiotic spectrometer delivers $2\times10^5$~\si{n/s/cm^2} at 3\% elastic energy resolution and 150 Hz without any pulse-shaping chopper. Both REMORA and LET can provide multiple incident energies: REMORA transmits higher-order harmonics of the fundamental wavelength, but with reduced flux, whereas LET provides several incident energies at comparatively similar fluxes. LET is in constant high demand, a comparable symbiotic instrument at the ESS would be highly productive by performing experiments that are routine today, relieving pressure on the flagship instruments. Other instruments such as SHARPER at the ILL or FOCUS at SINQ, PSI, \cite{Zanotti2026TheSHARP,Mesot1996Focus:Institute} have similar energy resolution, but higher flux, primarily due to having an entire guide to themselves. As an added benefit of the ESS being a pulsed source and having a high brilliance, the flux at $(004)$ remains high and usable, at $7\times10^4$~\si{n/s/cm^2}, sufficient for much of the science being done today. 

In terms of background, the conditions for a crystal monochromator direct spectrometer at the ESS are excellent. On a guide transporting $\lambda < \qty{1.5}{\angstrom}$ neutrons, line-of-sight can be lost upstream the monochromator position, and thus the secondary fast neutrons propagating in the guide shielding enclosure will not illuminate the spectrometer tank directly. Small amounts of concrete shielding enclosure is necessary as the beam is inherently low flux, and larger amounts of fast neutron background shielding can be strategically placed around the spectrometer tank, similar to that of SHARPER at ILL. 

Several practical considerations decide which flagship instruments make suitable hosts. The first is monochromator insertion loss. Extracting flux via Bragg reflection inevitably perturbs the transmitted beam, and this perturbation must remain within acceptable bounds for the host, never compromising the high flux of the flagship instruments. With HOPG, the (002) reflection accounts for the dominant loss, while higher-order contributions are substantially smaller \cite{Freund2022OptimizationOptics,Moore2000NeutronGraphite, Adib2012CharacteristicsMonochromator}; careful choice of monochromator geometry and extraction angle, can limit the impact to a level the host instrument's flux budget can absorb. Crystal material also plays a role, alternatives such as Si offer lower beam attenuation where reduced perturbation to the host is a priority. All of this needs to be rigorously simulated and tested before implementing a REMORA-like instrument on a beamline. Similarly, if the flagship instrument employs a tightly restrictive chopper configuration upstream of the extraction point, the transmitted bandwidth at the monochromator may become insufficient to support higher-order harmonics, leaving out the multi-energy advantage demonstrated above. Host instruments that routinely operate with strongly time-structured beams may therefore not consistently provide the spectral range required for stable symbiotic performance, and compatibility must be assessed on a case-by-case basis.

The second criterion concerns the beam state at the extraction point, both spectrally and in angular phase space. Instruments with early pulse-shaping choppers, such as BIFROST, may in high-resolution configurations have already discarded much of the available bandwidth before extraction \cite{Toft-Petersen2025BIFROSTAnSource}. By contrast, instruments such as T-REX, with frame-definition choppers further downstream, preserve a broader beam \cite{Voigt2014ChopperSources}. The proposed NJORD concept similarly leaves the beam comparatively open for symbiotic extraction \cite{Fogh2026ProposalRemora}. Host suitability therefore depends not only on flux, but on the spectral and angular phase space remaining available and stable at the extraction point.

Instruments such as REMORA would perform similarly to existing cold ToF spectrometers in the energy transfer regime 0-20 meV, which are already in very high demand. The science cases range from fundamental magnetism, like loop liquids in pyrochlores \cite{Hemmatzade} and quasielastic dynamics in skyrmionic systems \cite{Nikita} to protein dynamics \cite{Protein}. A REMORA-like instrument is expected to cost approximately half as much as a flagship instrument such as T-REX, since the primary optics, neutron guide, choppers, and shielding are already in place, leaving the secondary spectrometer and its large detector bank as the main additional costs.

The symbiotic principle demonstrated here is not tied to the specific flat HOPG arrangement used for REMORA. A doubly focusing, spinning monochromator of the kind described in \cite{Zanatta2019} could instead focus the extracted beam onto a smaller footprint and ease host-compatibility constraints. Symbiotic extraction is not limited to spectrometers: diffraction, imaging, or reflectometry stations could equally share a guide with a flagship host. REMORA should therefore be understood as one concrete realization of a broader design space, with the extraction optic and secondary instrument type matched to the beam conditions available at a given host.


\section{Methodological Discussion}
The gradient-descent optimization converged efficiently for both designs, though with notable differences in character. For the stacked diffractometer, the FOM \cite{Cussen2016ComputerConfigurations} weights high-angle ($2\theta> 90$) resolution heavily through the $\sin(\theta)$ factor embedded in FOM, biasing the optimization toward flat resolution curves and good backscattering performance, as reflected in the results above. For the symbiotic spectrometer, a two-step strategy was used instead: resolution was first constrained to meet the target threshold, and flux subsequently maximized within that constraint. This decouples resolution design from intensity optimization and removes the need for an ad-hoc scalar FOM, at the cost of a somewhat more involved optimization procedure.

Several limitations of the current approach should be acknowledged. Monte Carlo statistical noise introduces fluctuations of order 10\% in the FOM during optimization loops, and strongly coupled parameters, most notably the inter-element distances, were treated through sequential one-dimensional scans rather than joint exploration. A more comprehensive campaign using Bayesian optimization or machine-learning-driven methods would likely reveal a more robust global optimum and expose deeper physical coupling between parameters. These refinements are left to future work; the designs presented here are sufficient to establish the scientific and technical feasibility of both concepts.

\section{Conclusions \& Outlook}
\label{sec:outlook}
We have demonstrated the feasibility and good performance of two symbiotic instrument concepts at the ESS: a diffraction concept that can be scaled at relatively low cost and a spectrometer requiring investments primarily in the secondary spectrometer. The simulated performance is on par with highly successful instruments on both existing and recently decommissioned neutron sources. Neutron scattering serves a diverse neutron community, and not all experiments require world-leading performance. If such instruments can be built without compromising on the instrument suite in general, they offer a means of expanding capacity by utilizing existing infrastructure, and broadening access to neutron beamtime with limited investment in addition to the beamline components themselves. 


More broadly, additional capacity is urgently needed, while all known neutron production methods for material science require large investments in infrastructure, maintenance and operation. Maximizing the scientific return from existing investments is not only prudent but crucial.. The concepts outlined represent two potential pathways. Many variations of these symbiotic concepts could exploit the unique specifications of the ESS, enabling complementary instruments and providing much-needed additional beamtime capacity.

     
\appendix 

%
%
%
%
%
\section{Optimized Instrumental Parameters}
\subsection{SLEIPNIR Powder Diffractometer}
\label{sec:SLEIPNIR_params}

The optimized parameters for the SLEIPNIR powder diffractometer leg are
summarized in \cref{tab:SLEIPNIR_params}. The full McStas instrument file
is available at
\href{https://github.com/nicolai3008/SLEIPNIR}{github.com/nicolai3008/SLEIPNIR}.

The elliptical guide described in \cref{tab:SLEIPNIR_params} was not
subject to optimization and serves primarily to transport flux from the
moderator to the first virtual source. Its parameters therefore represent
a reasonable starting point rather than a converged solution, and further
optimization would likely yield a modest increase in flux at the sample.

An alternative optical approach, not pursued in the present study, would
be to replace the elliptical guide with a parabolic guide that produces a
low-divergence beam at the monochromator position. In this configuration,
the first monochromator would operate in plane-to-point Rowland geometry
rather than the point-to-point scheme used here, focusing directly from
the monochromator face to the second virtual source. This would trade some
flux for improved wavelength resolution at the first monochromatization
stage and may be better suited to the use of multiple legs, since the
monochromator size has to scale with the distance from the first virtual
source when using an elliptical guide.

\begin{table}[ht]
\caption{Optimized parameters for the SLEIPNIR powder diffractometer. The
elliptical guide has focal points at the moderator and \SI{1.5}{m} before
the first monochromator, denoted Virtual Source 1. The virtual source
between the two monochromators is denoted Virtual Source 2.}
\label{tab:SLEIPNIR_params}
\begin{center}
\begin{tabular}{ll}
\hline
Parameter & Value \\
\hline
Fundamental wavelength & \SI{2.6}{\angstrom} \\
Elliptical guide length & \SI{10}{m} \\
Elliptical guide $m$-value & 2 \\
Virtual Source 1 to monochromator 1, $L_1$ & \SI{1.5}{m} \\
Monochromator 1 mosaicity & \SI{30}{arcmin} \\
Monochromator 1 height $\times$ width & \SI{12}{cm} $\times$ \SI{9}{cm} \\
Monochromator 1 to Virtual Source 2, $L_1$ & \SI{1.5}{m} \\
Virtual Source 2 to monochromator 2, $L_2$ & \SI{3.0}{m} \\
Monochromator 2 mosaicity & \SI{80}{arcmin} \\
Monochromator 2 height $\times$ width & \SI{12}{cm} $\times$ \SI{9}{cm} \\
Monochromator 2 to sample, $L_2$ & \SI{3.0}{m} \\
Instrument height, $H$ & \SI{3.2}{m} \\
Horizontal collimation & \SI{10}{arcmin} \\
Vertical collimation & \SI{90}{arcmin} \\
Detector bank radius, $L_3$ & \SI{0.75}{m} \\
\hline
\end{tabular}
\end{center}
\end{table}

\subsection{REMORA}\label{sec:REMORA_params}

The optimized parameters for the REMORA symbiotic spectrometer are summarized in \cref{tab:REMORA_params}. The full McStas instrument file is available at
\href{https://github.com/nicolai3008/Njord_Remora/blob/main/Remora/remora.md}{\texttt{https://github.com/nicolai3008/REMORA}}. 

As with SLEIPNIR, the guide section used here is not optimized for symbiotic operation: the T-REX guide is used as-is up to the end of its curved section, after which a simple $m=2$ straight section transports the beam to the monochromator. The monochromator dimensions are constrained by the T-REX guide aperture rather than by the optimization. A dedicated extraction guide, shaped to match the monochromator acceptance and minimize divergence at the crystal face, would likely yield a meaningful flux improvement and is a natural target for future design work. 

A simplified variant of this design was also explored, in which the NMO, collimator and guide section between the monochromator and the focusing optics were removed entirely. In this configuration, beam collimation is achieved through distance alone, and the monochromator is operated in doubly focusing geometry to compensate for the absence of the NMO. This variant represents the minimal-cost implementation of the concept and serves as a lower bound on instrument complexity, at the expense of the instrument being less compact.

\begin{table}[ht]
\caption{Optimized parameters for the REMORA symbiotic spectrometer.}
\label{tab:REMORA_params}
\begin{center}
\begin{tabular}{ll}
\toprule
Parameter & Value \\
\midrule
Fundamental wavelength & \SI{4.8}{\angstrom} \\
Moderator to monochromator, $d_{\mathrm{sm}}$ & \SI{66}{m} \\
Monochromator mosaicity & \SI{30}{arcmin} \\
Monochromator height $\times$ width & \SI{9}{cm} $\times$ \SI{6}{cm} \\
Monochromator to NMO guide $m$-value & 2 \\
Horizontal Collimation & \SI{30}{arcmin}\\
NMO relative position, $r_{\mathrm{NMO}}$ & $0.5 \cdot d_{\mathrm{ms}}$ \\
NMO length, $l_{\mathrm{NMO}}$ & \SI{0.3}{m} \\
Fermi chopper relative position, $r_{\mathrm{fc}}$ & $0.85 \cdot d_{\mathrm{ms}}$ \\
Fermi chopper slit width & \SI{1.2}{cm} \\
Monochromator to sample, $d_{\mathrm{ms}}$ & \SI{1.3}{m} \\
Sample to detector, $d_{\mathrm{sd}}$ & \SI{2.52}{m} \\
\bottomrule
\end{tabular}
\end{center}
\end{table}

\begin{acknowledgements}
The authors thank the European Spallation Source ERIC for supporting the development of the REMORA concept.
\end{acknowledgements}

\begin{funding}
This work was supported by the Data Management and Scientific Computing Division (DMSC) of the European Spallation Source ERIC through support for the development of the REMORA concept.
\end{funding}

\ConflictsOfInterest{The authors declare that there are no conflicts of interest.}

\DataAvailability{The models developed for this work are available from the authors' \href{https://github.com/nicolai3008}{GitHub repository}. The experimental and simulation data supporting the results presented in this article are available from the corresponding author upon reasonable request.}

\bibliography{fixed_bibliography} 

\end{document}